\documentclass{article}

\usepackage{algorithm}
\usepackage{algpseudocode}
\usepackage{amsfonts}
\usepackage{amsmath}
\usepackage{arxiv}
\usepackage{authblk}
\usepackage{booktabs}
\usepackage{color, colortbl}
\usepackage[T1]{fontenc}
\usepackage{hyperref}
\usepackage{graphicx}
\usepackage[utf8]{inputenc}
\usepackage{lipsum}
\usepackage{microtype}
\usepackage{multirow}
\usepackage{nicefrac}
\usepackage{subfigure}
\usepackage[table,xcdraw]{xcolor}
\usepackage{xr}
\usepackage{url}

\usepackage{booktabs} 
\usepackage{threeparttable}
\usepackage{rotating}
\usepackage{pdflscape}

\graphicspath{ {./images/} }

\title{Meta-Path-Free Representation Learning on Heterogeneous Networks}

\author[1,2]{Jie Zhang}
\author[1]{Jinru Ding} 
\author[1]{Suyuan Liu}
\author[ 3]{Hongyan Wu}

\affil[1]{SenseTime Research, Shanghai, China.} 
\affil[2]{Qing Yuan Research Institute, Shanghai Jiao Tong University, Shanghai,  China.}  

\affil[3]{Shenzhen Institutes of Advanced Technology,  Chinese Academy of Sciences, Shenzhen, China.} 

\begin{document}
\maketitle	

\begin{abstract}

Real-world networks and knowledge graphs are usually heterogeneous networks. Representation learning on heterogeneous networks is not only a popular but a pragmatic research field. The main challenge comes from the heterogeneity---the diverse types of nodes and edges. Besides, for a given node in a HIN, the significance of a neighborhood node depends not only on the structural distance but semantics. How to effectively capture both structural and semantic relations is another challenge. The current state-of-the-art methods are based on the algorithm of meta-path and therefore have a serious disadvantage---the performance depends on the arbitrary choosing of meta-path(s). However, the selection of meta-path(s) is experience-based and time-consuming. In this work, we propose a novel meta-path-free representation learning on heterogeneous networks, namely Heterogeneous graph Convolutional Networks (HCN). The proposed method fuses the heterogeneity and develops a $k$-strata algorithm ($k$ is an integer) to capture the $k$-hop structural and semantic information in heterogeneous networks. To the best of our knowledge, this is the first attempt to break out of the confinement of meta-paths for representation learning on heterogeneous networks. We carry out extensive experiments on three real-world heterogeneous networks. The experimental results demonstrate that the proposed method significantly outperforms the current state-of-the-art methods in a variety of analytic tasks.

\end{abstract}


\section{Introduction}

\begin{figure}[h]
	\centering 
	\includegraphics[width=0.5\linewidth]{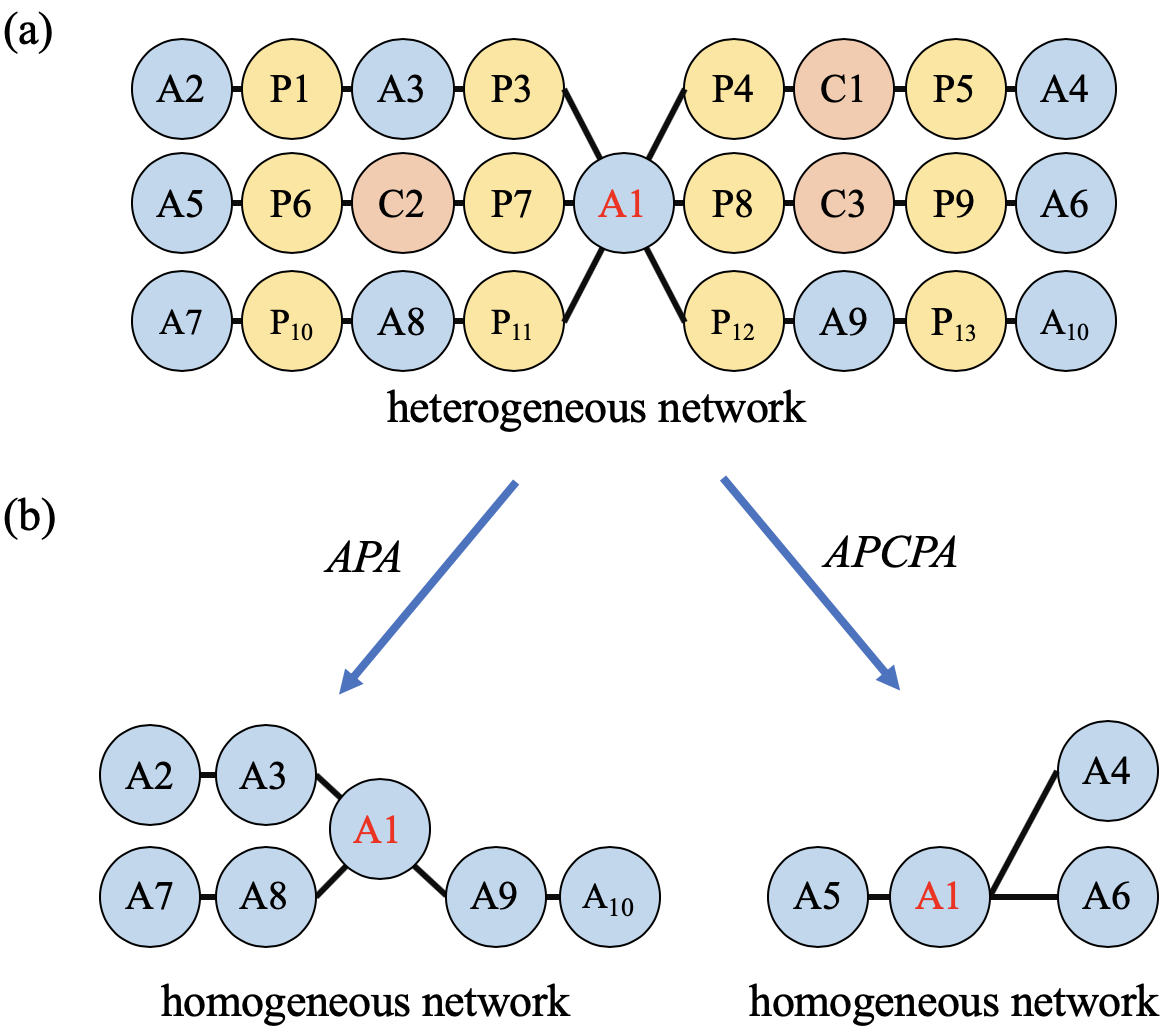}
	\caption{(a) An illustrative example of a DBLP-like network. The ``A"s, ``P"s, and ``C"s represent nodes of authors, papers, and conferences, respectively. The ``A1" is the given node. (b) An illustrative example of the pretreatment of HAN. By $APA$ and $APCPA$, the heterogeneous network is decomposed and reorganized into two sub-homogeneous networks. The neighborhood in the two new sub-homogeneous networks is the meta-path-based ($APA$ and $APCPA$) neighborhood in the original heterogeneous network. }
	\label{fig:HIN-HAN}
\end{figure}

Heterogeneous information networks (HIN) is a type of networks that involve multiple types of nodes and/or edges \cite{sun2013mining}. Take Digital Bibliographic Library Browser (DBLP)\footnote{https://dblp.uni-trier.de} as an example. The node types include authors ($A$), papers ($P$), and conferences ($C$). And the edge types include a writing relation between a paper ($P$) and an author ($A$) and a publishing relation between a paper ($P$) and a conference ($C$). Figure \ref{fig:HIN-HAN}(a) gives an example of DBLP-like networks.

Real-world networks are usually HINs. For instance, publication networks \cite{giles2006future}, biological networks \cite{Roy2007Integrative}, highway networks \cite{Jiang2005Knowledge}, and most knowledge graphs are HINs. Representation learning on HINs, also known as heterogeneous network embedding (HNE), captures semantic and structural information by embedding diverse types of nodes and/or the entire network into a low-dimensional space. HNE effectively helps downstream analytical tasks, such as knowledge-guided recommendation systems \cite{park2018collaborative, wang2019multi}, knowledge-based image classification \cite{zhang2019knowledge, marino2016more} and captioning \cite{zhou2019improving, li2019knowledge}, knowledge-guided natural language processing (NLP) \cite{zhang2019ernie, yao2019clinical}, and so on. Therefore, representation learning on HINs is not only a popular but also a pragmatic research field.

There are mainly two challenges for HNE. \textbf{[Challenge 1]} A heterogeneous network has much more complicated semantics than a homogeneous network. Diverse types of nodes and edges have various feature spaces and semantic meaning. The challenge of the heterogeneity cannot be simply handled by the methods of homogeneous network embedding \cite{chang2015heterogeneous, shi2018easing, wang2019heterogeneous}. \textbf{[Challenge 2]} For a given node in a HIN, the significance of a neighborhood node depends not only on the structural distance but semantics.  \cite{chen2018pme}. In other words, a farther neighbor may have more significance. Some researches \cite{fu2017hin2vec, wang2019heterogeneous} on HNE find the analytical outcomes based on a long-distance neighborhood outperform those based on a short-distance neighborhood in node clustering tasks on DBLP.

Classic algorithms for HNE, such as Metapath2vec, apply the algorithm of meta-path.  A meta-path is a pre-defined sequence of node types. Metapath2vec takes the meta-path-guided random walks and then applies a skip-gram algorithm \cite{dong2017metapath2vec}. Recently, Graph Neural Networks (GNN), such as Graph Convolutional Networks (GCN) \cite{bruna2013spectral, defferrard2016convolutional, kipf2016semi} and Graph Attention Networks (GAT) \cite{velivckovic2017graph}, have shown superior performance on homogeneous network embedding. Therefore, the current state-of-the-art methods, such as Heterogeneous graph Attention Network (HAN) \cite{wang2019heterogeneous}, combines the algorithms of meta-path and GNN to perform HNE. 

However, there is a serious disadvantage of meta-path-based methods---the meta-paths are either specified by users or derived from supervision \cite{dong2017metapath2vec, fu2017hin2vec, shi2018easing, wang2019heterogeneous}. The meta-paths selected in these ways only reflect certain aspects of HIN \cite{shi2018easing}, and different meta-paths result in different outcomes. Researchers need to explore meta-paths as many as possible and choose the best meta-path(s) \cite{fu2017hin2vec}. However, the number of meta-paths is infinite and researchers can hardly test all possible meta-paths. Therefore, the selection of meta-paths is usually experience-based and time-consuming \cite{chen2017task, li2017semi, shang2016meta}.

We develop a novel meta-path-free representation learning on heterogeneous networks, namely Heterogeneous graph Convolutional Networks (HCN). The proposed method develops a meta-path-free $k$-strata algorithm, which naturally incorporates miscellaneous composite relations in heterogeneous networks. The hybrid of miscellaneous composite relations is the key to fusing the heterogeneity and capturing both structural and semantic information in heterogeneous networks without arbitrarily selecting meta-paths. 

Furthermore, the pretreatment of the proposed method is much easier. Comparatively, for the pretreatment of meta-path-based methods,  such as the pretreatment of HAN in Figure \ref{fig:HIN-HAN}(b), the time complexity depends on the length of the meta-paths, the number of nodes and branches in HIN, and the number of meta-paths. 


The contributions of this work are as follows: (1) to the best of our knowledge, this is the first attempt to break out of the confinement of meta-paths for HNE; (2) the proposed method can capture both the semantic and structural relations; and (3) we carry out extensive experiments on three real-world HINs, and the results prove that the proposed method significantly outperforms the current state-of-the-art methods in a variety of analytical tasks.

\begin{figure}[h]
	\centering 
	\includegraphics[width=0.85\linewidth]{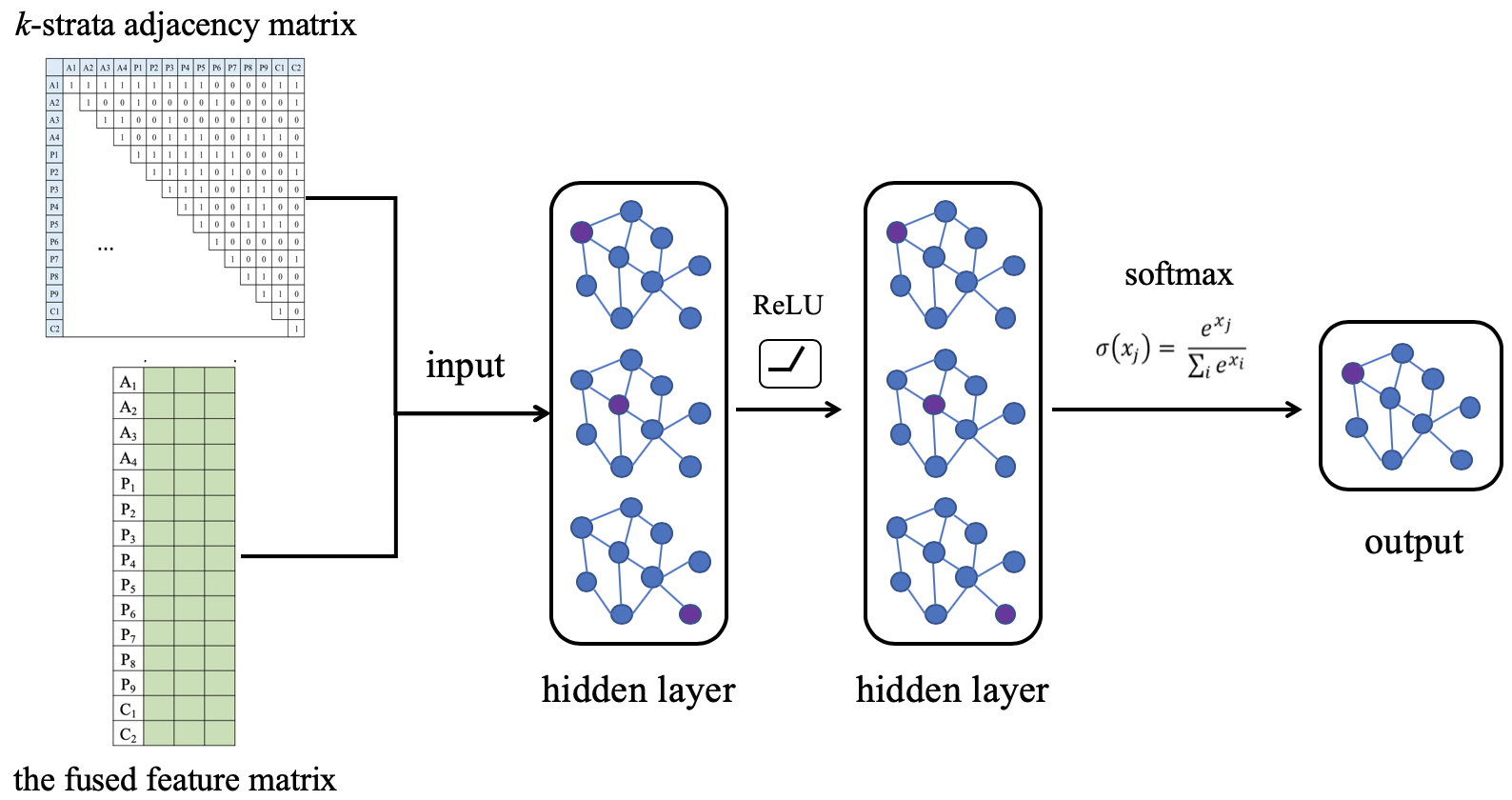}
	\caption{The representation learning implemented in a two-layered GNN. The inputs are the $k$-strata adjacency matrix and the fused feature matrix. The output is analytical outcome of a node classification task.}
	\label{fig:structure}
\end{figure}


\section{RELATED WORK}

Metapath2vec is a meta-path-based unsupervised learning. There are two terms in Metapath2vec: (1) the meta-path scheme and (2) the meta-path instance. (1) A meta-path scheme is a pre-defined sequence of node types. Take DBLP as an example. The commonly used meta-path schemes for DBLP are ``author-paper-author" ($APA$) and ``author-paper-conference-paper-author" ($APCPA$). (2) A meta-path instance is a node sequence that follows and repeats the format of a meta-path scheme until it reaches a fixed length, which is set to 100 in Metapath2vec. Take Figure \ref{fig:HIN-HAN}(a) as an example. By following and repeating the meta-path scheme of $APA$, meta-path instances, such as $A_{2}P_{1}A_{3}P_{3}A_{1}P_{12}A_{9}P_{13}A_{10}$, are generated. The generated meta-path instances are inputted into the skip-gram algorithm to learn HNE \cite{dong2017metapath2vec}. Please note that all meta-path instances are ``randomly" generated. The ``randomness" may generate some meta-path instances but neglect others.

HIN2Vec is a meta-path-based supervised learning. HIN2Vec explores all meta-path instances within $w$ hops and performs link predictions to achieve HNE. The HIN2Vec compares the length of hops $w$ in four HINs: Blogcatalog\footnote{http://socialcomputing.asu.edu/datasets/BlogCatalog3}, Yelp\footnote{https://dblp.uni-trier.de}, U.S. Patents\footnote{http://www.dev.patentsview.org/workshop/ participants.html}, and DBLP, and finds that a longer meta-path instances is crucial for a complicated HIN, such as DBLP, because a longer meta-path may have a significant semantic meaning \cite{fu2017hin2vec}, such as $APAPA$---two authors have co-authorship with the same author. 

HAN is a typical algorithm that combines the algorithms of meta-path and GNN. The analytical process is divided into three steps. Firstly, by pre-defined meta-paths, a HIN is decomposed and reorganized into several homogeneous networks. Take Figure \ref{fig:HIN-HAN} as an example. By $APA$ and $APCPA$, the heterogeneous network in Figure \ref{fig:HIN-HAN}(a) is decomposed and reorganized into two sub-homogeneous networks in Figure \ref{fig:HIN-HAN}(b). The neighborhood in the sub-homogeneous networks is the meta-path-based ($APA$ and $APCPA$) neighborhood in the original heterogeneous network. Secondly, HAN leverages the GNN algorithm to learn node embedding in the two new sub-homogeneous networks. Thirdly, the two pieces of node embedding learned from the two new sub-homogeneous networks are fused. Different from Metapath2vec and HIN2Vec, HAN achieves embedding of only one type of node. 

In summary, the ``randomness" in Metapath2vec might neglect some meta-path instances and thereby lose some indispensable information. HIN2Vec finds that the length of meta-paths impacts the analytical performance, but unfortunately, does not give a method to avoid choosing meta-paths. HAN achieves embedding of only one type of node in HINs. The selection of meta-paths in Metapath2vec, HIN2Vec, and HAN all strongly depends on the task at hand. 


\section{PRELIMINARY}

This section formally defines  (1) HIN and (2) distance between two nodes. Table \ref{tab:HIN} presents the notations in this work.



\textit{Definition 3.1 } \textbf{Heterogeneous Information Network (HIN) \cite{sun2013mining}}. A HIN, denoted as $G = (V, E)$, is composed of a set of nodes $V$ and a set of edges $E$. And $O$ and $R$ denote the set of node types and edge types, respectively, and $|O| + |R| > 2$. 


\begin{table*}

	\centering
	
	\caption{Notations and Explanations.}
	
	\label{tab:notation}
	
	\begin{tabular}{cc}
		
		\toprule
		
		Notations & Explanations\\
		
		\midrule
		
		$G$ & a heterogeneous network \\
		
		$V$ & set of all nodes\\
		
		$E$ & set of all edges\\
		
		$O$ & set of all node types\\
		
		$R$ & set of all edge types\\
		
		$\tilde{\mathcal{A}}^k$ & $k$-strata adjacency matrix\\
		
		$M$ & type-specific transformation matrix\\
		
		$X'$ & fused feature matrix\\
		
		$Z$ & final embedding\\
		
		\bottomrule
		
	\end{tabular}
	
\end{table*}





\textit{Definition 3.2 } \textbf{Distance between Two Nodes}.The $distance(i, j)$ is the number of the hops in the shortest path between two given nodes $i$ and $j$. Especially, the distance from a node to itself is 0; and the distance is infinite ($\infty$) if no path exists between $i$ and $j$. The Formula (\ref{eq:dist}) illustrates the definition of the distance between two nodes.

\begin{equation}
	\label{eq:dist}
	distance(i, j)=\left\{
	\begin{array}{rcl}
		0   & {i = j}\\
		k   & {\mbox{$k$ hops in the shortest path between } i \mbox{ and } j}\\
		\infty   & {\mbox{no path between } i \mbox{ and } j}
	\end{array} \right.
\end{equation}

%


\section{META-PATH-FREE REPRESENTATION LEARNING}

This section explains the proposed meta-path-free representation learning for HNE.

\subsection{$k$-Strata}

\begin{figure}[h]
	\centering 
	\includegraphics[width=0.705\linewidth]{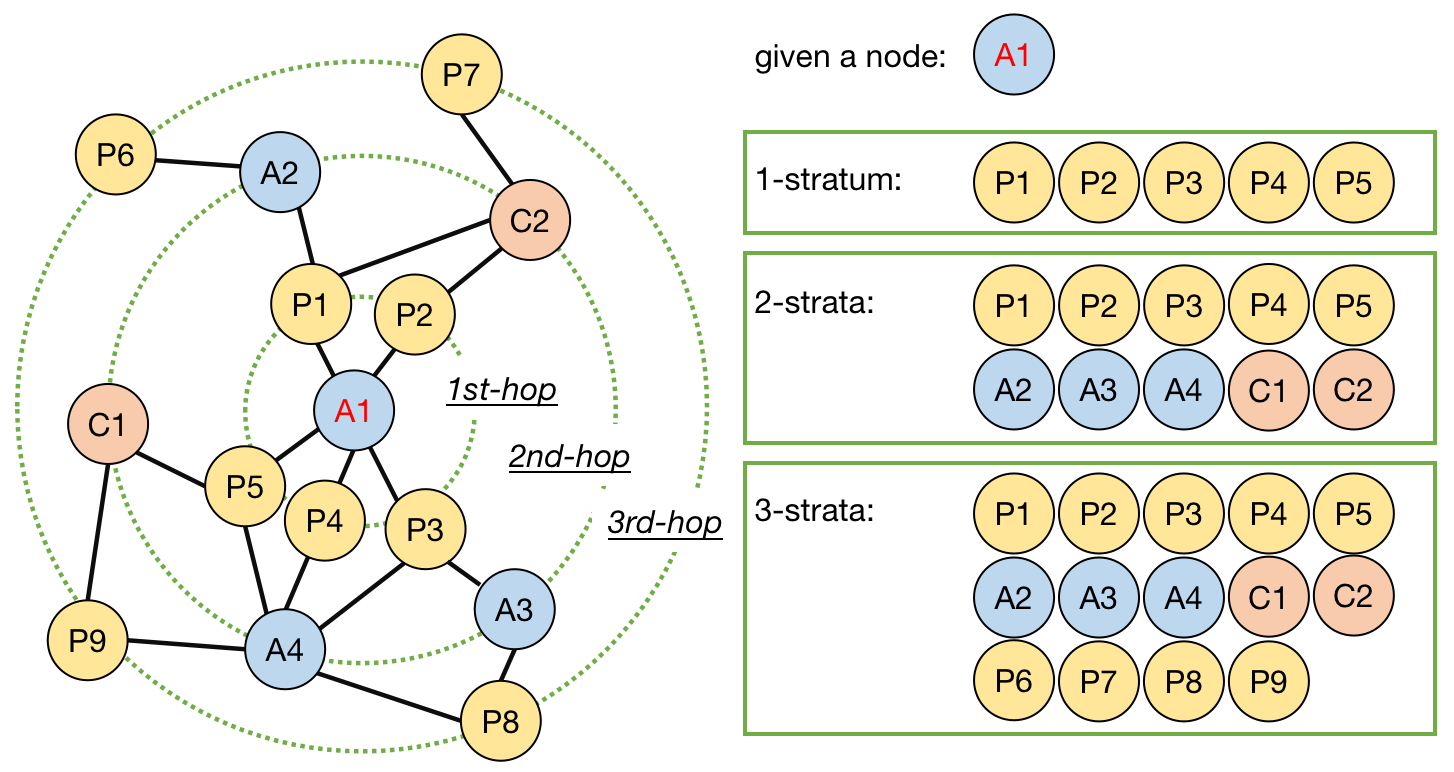}
	\caption{An example of the algorithm of $k$-strata in a DBLP-like graph. The ``A"s, ``P"s, and ``C"s represent nodes of authors, papers, and conferences, respectively. The ``A1" is the given node. The $k$-strata refers to all the nodes within the k-hop range from the given node $A1$. For example, the 2-strata of $A1$ includes $P1$, $P2$, $P3$, $P4$, $P5$, $A2$, $A3$, $A4$, $C1$, and $C2$. For a given node in a HIN, the significance of a neighborhood node depends not only on the structural distance but semantics. For example, $A1$ published two papers ($P1$ and $P2$) in a conference ($C2$). The $P1$ is a paper that introduces how to use knowledge graph embedding to enrich word embedding dimensions; the $P2$ is a paper that performs reinforcement learning in a question and answer (QA) system, and the $C2$ is the conference of Association for Computational Linguistics (ACL). The $C2$ reflects the $A1$'s research area and interests (computational linguistic) more directly and obviously than $P1$ or $P2$. }
	\label{fig:k_strata}
\end{figure}

\begin{figure}[h]
	\centering 
	\includegraphics[width=0.6\linewidth]{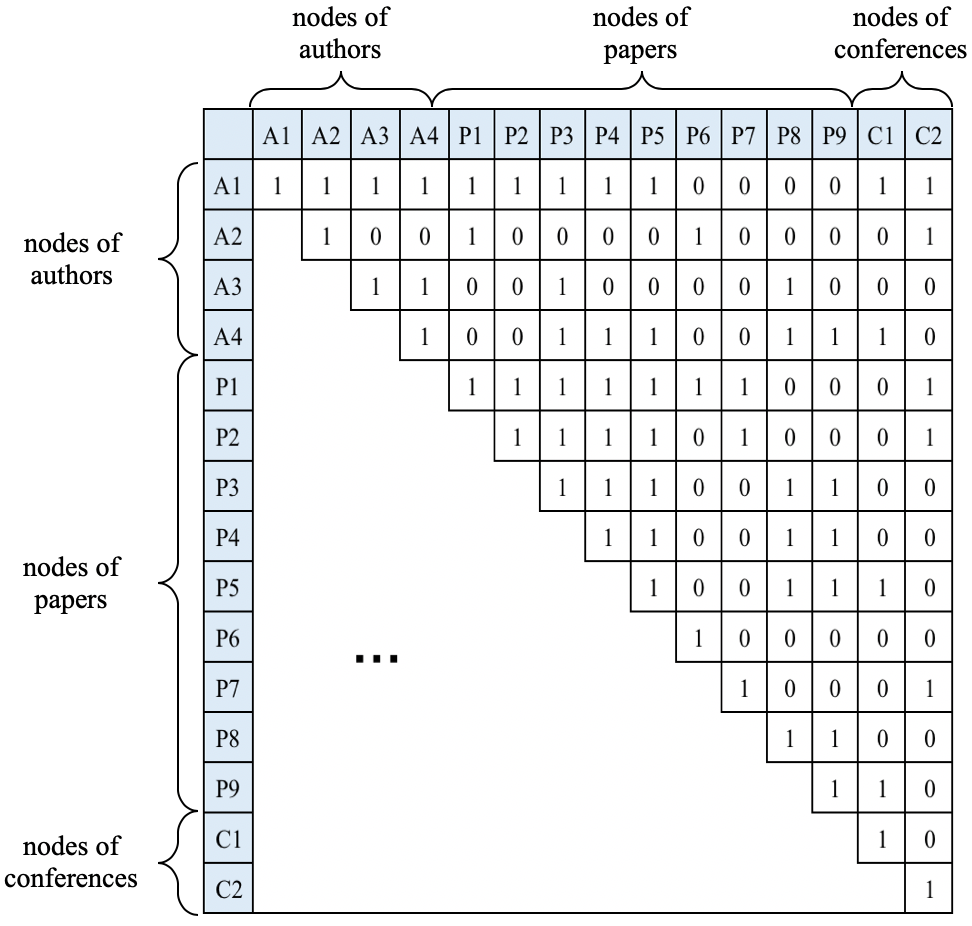}
	\caption{An illustration of a $k$-strata adjacency matrix. The two-strata adjacency matrix in this figure corresponds to the DBLP-like network in Figure \ref{fig:k_strata}. Since a $k$-strata adjacency matrix is a symmetric matrix, this figure only shows the upper right half.}
	\label{fig:k_strata_adj}
\end{figure}


For a given node in a homogeneous network, the significance of a neighborhood node is more relevant to the structural distance. The longer the distance is, the less the significance could be. Comparatively, for a given node in a HIN, the significance of a neighborhood node depends not only on the structural distance but semantics. Take Figure \ref{fig:k_strata} as an example. The $A1$ published two papers ($P1$ and $P2$) in a conference ($C2$). The $P1$ is a paper that introduces how to use knowledge graph embedding to enrich word embedding dimensions; the $P2$ is a paper that performs reinforcement learning in a question and answer (QA) system; and the $C2$ is the conference of Association for Computational Linguistics (ACL). Thereby, $C2$ reflects the $A1$'s research area and interests (computational linguistic) more directly and obviously than $P1$ or $P2$. Another example is also from Figure \ref{fig:k_strata}. The $A4$ and $A1$ from the same lab share the same research area interests. The $A4$ and $A1$ have three co-author papers $P3$, $P4$, and $P5$, and the $P3$, $P4$, and $P5$ use different algorithms. Therefore, for the given node $A1$, the $A4$ has more significance than $P3$, $P4$, or $P5$, although $A4$ is farther than $P3$, $P4$, or $P5$. In conclusion, when performing representation learning in a HIN, we need to take both distance and semantics into consideration. 

We introduce the concept of $k$-strata ($k$ is an integer), to refer to all nodes within the k-hop range from a given node, as Figure \ref{fig:k_strata} illustrates. Formula (\ref{eq:$k$-strata}) illustrates the $k$-strata adjacency matrix $\tilde{\mathcal{A}}^k$. The value $\tilde{\mathcal{A}^k_{i, j}}$ between two nodes $i$ and $j$ is defined as:

\begin{equation}
	\label{eq:$k$-strata}
	\tilde{\mathcal{A}^k_{i, j}}=\left\{
	\begin{array}{rcl}
		1 & { distance(i, j) \leq k}\\
		0 & { distance(i, j) > k}
	\end{array} \right.
\end{equation}

where $\tilde{\mathcal{A}}^k \in \mathbb{R}^{n \times n}$ ; $n$ is the number of all nodes. Please note that the $\tilde{\mathcal{A}}^k$ includes self-connections since the distance from a node to itself is 0. Figure \ref{fig:k_strata_adj} is a two-strata adjacency matrix, which corresponds to the Figure \ref{fig:k_strata}. Since the $k$-strata adjacency matrix is a symmetric matrix, Figure \ref{fig:k_strata_adj}  only shows the upper right half. Please note that the two-strata adjacency matrix considers all the  relations between any two nodes within two-hop range. 

Algorithm \ref{alg:$k$-strata} explains how to generate the $k$-strata adjacency matrix. Although it might look complicated, the implementation could be quite simple, just one line in the Pandas\footnote{https://pandas.pydata.org/} code as follows:

$\tilde{\mathcal{A}}^k \leftarrow \tilde{\mathcal{A}}^{k-1}\mbox{.apply(lambda x: (}\tilde{\mathcal{A}^1}\mbox{[x==1].any()).astype(int))}$


\begin{algorithm}[htb]
	\caption{The generation of the $k$-strata adjacency matrix.}
	\label{alg:$k$-strata}
	\begin{algorithmic}[1]
		\Require 
		The heterogeneous graph $G=(V, E)$, \\
		The 1-stratum adjacency matrix $\tilde{\mathcal{A}^1}$,\\
		The number of strata $K$ ($K \ge 2$).
		\Ensure 
		The $k$-strata adjacency matrix $\tilde{\mathcal{A}}^k$.
		\For{$k = 2 $ ... $K$}
		\For{$i \in V$}
		\State From $\tilde{\mathcal{A}}^{(k-1)}$, find all the $(k-1)$-strata neighbors of the node $i$, denoted as a set $N_{i}^{(k-1)}$;
		\State From $\tilde{\mathcal{A}}^{1}$, find all the 1-hop neighbors of all the nodes in the set $N_{i}^{(k-1)}$, denoted as a set $N_i^{k-th}$; \\
		\State $N_{i}^{(k)} \leftarrow$ logical\_or ($N_{i}^{(k-1)}$, $N_i^{k-th}$); \\
		\State Append $N_i^{k}$ to matrix $\tilde{\mathcal{A}}^k$; \\
		\EndFor 
		\EndFor 
		\State \textbf{return} $\tilde{\mathcal{A}}^k$ .
	\end{algorithmic}
\end{algorithm}

\textbf{$\bullet$ Composite Relations}

We introduce the concept of ``composite relations", which can help us understand the reason why the $k$-strata captures both structural and semantic information in heterogeneous networks. 
In a $k$-hop structure: $V_1 \stackrel{R_1}{\longrightarrow} V_2 \stackrel{R_2}{\longrightarrow} \cdots \stackrel{R_{k}} {\longrightarrow} V_{k+1}$, where $V_i$ are nodes and $R_i$ are one-hop edges (or simple relations). The $k$-hop relation ($R$) between node $V_1$ and $V_{k+1}$ can be formulated as $R = R_1 \circ R_2 \circ \cdots \circ R_{k}$, where $\circ$ denotes the composition operator on relation. Therefore, a $k$-hop ($k \ge 2$) relation between two nodes implies a composite-relation with a distance of $k$. Consequently, the $k$-strata adjacency matrix incorporates miscellaneous composite relations and therefore has two advantages for learning HNE.

Firstly, the $k$-strata adjacency matrix captures composite relations between any two nodes, while meta-path-based methods only capture the relations along the meta-paths. Take the DBLP-like network in Figure \ref{fig:HIN-HAN} as an example. The Metapath2vec can only learn the relation of $APA$ based on the meta-path $APA$ (as shown on the left side of Figure \ref{fig:k_strata_adj}(b)),  and  the relation of $APCPA$ based on the meta-path $APCPA$ (as shown on the right side of  Figure \ref{fig:k_strata_adj}(b)). Although HAN can fuse several meta-paths, such as $APCPA$ and $APA$, in a real-world HIN, there could be much more useful composite relations and HAN cannot cover all of them. 

Secondly, the $k$-strata adjacency matrix can capture composite relations across meta-paths.    
Still take the DBLP-like network in Figure \ref{fig:HIN-HAN} (b)as an example. For a node classification task of authors, the relationship between $A6$ and $A3$ should be considered since both $A6$ and $A3$ have composite relations with $A1$.  However, in meta-path-based methods, the $A6$ can never capture semantic information from $A3$, since $A6$ and $A3$ are located in different meta-paths---$A3$ is in $APA$ while $A6$ is in $APCPA$.  For the proposed $k$-strata algorithm, the four-strata adjacency matrix includes a four-hop relation (between $A6$ and $A1$) and a two-hop relation (between $A1$ and $A3$). In other words, there is a consecutive relation of $A6-A1-A3$ in the four-strata adjacency matrix and $A6$ is $A3$'s neighbor's neighbor. Thereby, the $A6$ can capture semantic information from $A3$ through GNN.

In conclusion, the $k$-strata adjacency matrix incorporates miscellaneous composite relations. And the ``hybrid" of different composite relations is the key to fusing the heterogeneity and capturing both structural and semantic information in heterogeneous networks without arbitrarily selecting meta-paths. 



\subsection{Feature Fusion}
A HIN has different types of nodes and therefore has different feature spaces. For example, in a  DBLP-like network, the nodes of ``Author" have their own feature space. So do the nodes of ``Paper" and ``Conference". To achieve HNE, we need to fuse different feature spaces. 


\begin{figure}[h]
	\centering 
	\includegraphics[width=0.7\linewidth]{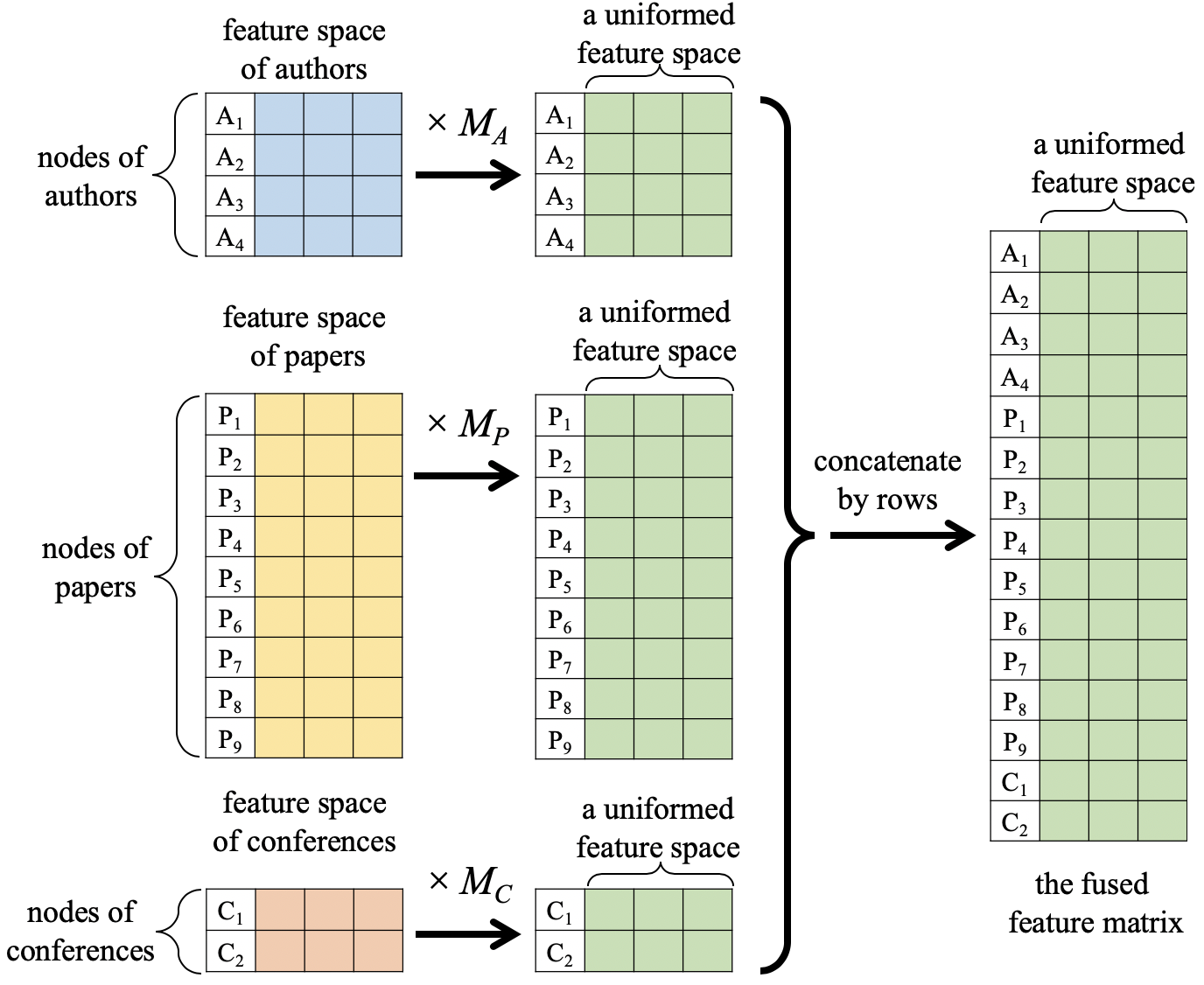}
	\caption{An illustration of the feature fusion. The original feature matrix of ``Author", ``Paper", and ``Conference" are multiplied by a type-specific transformation matrix $M_A$, $M_P$, and $M_C$, respectively, and then concatenated by rows. The example in this figure corresponds to the DBLP-like network in Figure \ref{fig:k_strata} and Figure \ref{fig:k_strata_adj}.}
	\label{fig:feature_fusion}
\end{figure}


\begin{algorithm}[htb]
	\caption{The algorithm of feature fusion.}
	\label{alg:ff}
	\begin{algorithmic}[1]
		\Require 
		the set of all node types $O$, \\
		the original feature matrix $X$, \\
		a trainable type-specific matrix $M_o$ for one node type $o$. \\
		\Ensure 
		The fused feature matrix $X'$.
		\For{$o \in O$}
		\State  From $X$, find the features of all the nodes of the type $o$, denoted as $X_o$
		\State $X'_o \leftarrow X_o M_o$ ; \\
		Append $X'_o$ to $X'$
		\EndFor 
		\State \textbf{return} $X'$.
	\end{algorithmic}
\end{algorithm}

We use a trainable type-specific transformation matrix for every node type and then append the transformed feature spaces. Figure \ref{fig:feature_fusion} illustrates the algorithm of feature fusion for the DBLP-like network in Figure \ref{fig:k_strata}. In particular, the original feature matrix of ``Author", ``Paper", and ``Conference" multiplies a trainable type-specific transformation matrix ($M_A$, $M_P$, and $M_C$), respectively. And then, we concatenate the transformed feature matrices of ``Author", ``Paper", and ``Conference" by rows. 

The fused feature matrix is denoted as $X' \in \mathbb{R}^{ n \times F'}$, where $n$ is the number of nodes and $F'$ is the dimension of the fused feature space. The Algorithm \ref{alg:ff} illustrates how to perform feature fusion for multiple types of nodes. 

The purpose of the trainable type-specific transformation matrix ($M_A$, $M_P$, and $M_C$) is to transform the different feature spaces into a unified feature space. 

\textbf{$\bullet$ Linear and non-linear transformation for feature fusion} 

For feature fusion in this work, we implement a linear transformation. One can also use a non-linear transformation by performing an activation function $\sigma$ after multiplying $M$, which amounts to a fully-connected layer. In other words, to achieve a non-linear transformation for feature fusion, one can implement one or more fully-connected layers. In this work, a linear transformation seems to perform well enough.

\subsection{Representation Learning}

To learn HNE, the $k$-strata adjacency matrix and the fused feature matrix are inputted to GNN, such as GCN or GAT, to perform a supervised node classification (Figure \ref{fig:structure}). In this work, we use GCN to implement representation learning. Formula (\ref{eq:forward}) shows the implementation. 

\begin{equation}
	\label{eq:forward}
	\begin{split}
		H^1 = \sigma(\hat{\mathcal{A}^k}X'W^{0}) \\
		H^2 = \sigma(\hat{\mathcal{A}^k}H^1W^{1}) \\
		... \\
		H^{h-1} = \sigma(\hat{\mathcal{A}^k}H^{h-2}W^{h-2}) \\
		Z = \hat{\mathcal{A}^k}H^{h-1}W^{h-1} \\
	\end{split}
\end{equation}

where $h$ is the number of GCN layers; $X' \in \mathbb{R}^{ n \times F'}$ is the fused feature matrix ; $W$ is a trainable weight matrix; $Z \in \mathbb{R}^{n \times C}$ is the final embedding matrix and $C$ is the dimension of the final embedding; $\sigma$ is an activation function and we use Rectified Linear Unit (ReLU); $\hat{\mathcal{A}^k} \in \mathbb{R}^{n \times n}$ is a symmetric normalized Laplacian $k$-strata adjacency matrix, defined as Formula (\ref{eq:hat_Ak}).

\begin{equation}
	\label{eq:hat_Ak}
	\hat{\mathcal{A}^k} = \tilde{D}^{-\frac{1}{2}}\tilde{\mathcal{A}}^k \tilde{D}^{-\frac{1}{2}}
\end{equation} 

where $\tilde{\mathcal{A}}^k$ is the $k$-strata-adjacency matrix and $\tilde{D}$ is the degree matrix , as shown in Formula (\ref{eq:tilde_D}) .

\begin{equation}
	\label{eq:tilde_D}
	\tilde{D}_{ii} = \sum_j\tilde{\mathcal{A}}^k_{ij} 
\end{equation}

For the multi-class classification, we calculate the cross-entropy loss over all labeled examples, as Formula (\ref{eq:loss}) shows.

\begin{equation}
	\label{eq:loss}
	\mathcal{L} = - \sum_{l \in \mathcal{Y}_L } Y_l \cdot \ln (softmax(Z_l) )
\end{equation}

where $\mathcal{Y}_L$ is a set of nodes that have labels; $Y_l \in \mathbb{R}^C$ is a vector indicating the true labels; $Z_l \in \mathbb{R}^C$ is the final embedding vector of a node that has a label; $softmax(Z_l)$ is the predicted probabilities of all classes, and $\cdot$ is dot product of two vectors.

\begin{algorithm}[htb]
	\caption{The representation learning implemented in GCN.}
	\label{alg:nmp_grl}
	\begin{algorithmic}[1]
		\Require
		The heterogeneous graph $G = \{V,E\}$, \\
		The fused feature matrix $X'$, \\
		The $k$-strata adjacency matrix: $\tilde{\mathcal{A}^k}$,\\
		The training epochs $T$.
		\Ensure
		The final embedding $Z$.
		\State $\tilde{D}_{ii} \leftarrow \sum_j\tilde{\mathcal{A}}^k_{ij} $;
		\State $\hat{\mathcal{A}^k} \leftarrow \tilde{D}^{-\frac{1}{2}}\tilde{\mathcal{A}}^k\tilde{D}^{-\frac{1}{2}} $;
		\For{$t \in T$}
		\State $Z \leftarrow GCN(\hat{\mathcal{A}^k}, X')$;
		\State Calculate loss: $\mathcal{L} = - \sum_{l \in \mathcal{Y}_L } Y_l \cdot \ln (softmax(Z_l) )$;
		\State Perform back propagation and update parameters;
		\EndFor 
		\State \textbf{return} $Z$.
	\end{algorithmic}
\end{algorithm}

The algorithm of the representation learning is described in Algorithm \ref{alg:nmp_grl}.

\subsection{Online Dilation}

A too large $k$ brings a too dense $k$-strata adjacency, which could increase training costs \cite{perozzi2014deepwalk}.     
A recent research theoretically demonstrates dropping edges reduces message passing in graph training \cite{rong2019truly}. To solve this problem, we randomly drop some $k$-strata edges, which we call ``dilation". In other words, the ``dilation" here means that we randomly choose a certain proportion of cells of ``1" in the $k$-strata adjacency matrix and change them to ``0" to make the $k$-strata adjacency matrix sparser. The proportion can be 30\%, 50\% or more. The operation is optional, and we can think of the dilation proportion as an adjustable hyper-parameter.  If such pretreatment of dilation does not bring worse analytical outcomes in heterogeneous networks, we can use the dilation to reduce training costs. 

In the implementation, to make the model robust, we adopt an ``online dilation" during model training, which performs a random drop every a few epochs. The algorithm of the representation learning with the ``online dilation" is described in Algorithm \ref{alg:nmp_grl_MSD}.

\begin{algorithm}
	
	\renewcommand{\algorithmicrequire}{\textbf{Input:}}
	
	\renewcommand{\algorithmicensure}{\textbf{Output:}}
	
	\caption{The representation learning with online dilation.}
	
	\label{alg:nmp_grl_MSD}
	
	\begin{algorithmic}[1]
		
		\Require
		
		the heterogeneous graph $G = \{V,E\}$, \\
		
		
		The fused feature matrix $X'$, \\
		
		The $k$-strata adjacency matrix: $\tilde{\mathcal{A}^k}$,\\
		
		The dilation proportion: $p \%$,\\
		
		The number $q$: perform a random drop every $q$ epochs,\\
		
		The training epochs $T$.
		
		\Ensure
		
		The final embedding $Z$.
		

		\For{$t \in T$}

		\If{$t$ mod $q == 0$} 
		
		
		\State $\tilde{\mathcal{A}}^{k-dilated} \leftarrow \mbox{randomly drop $p \%$ relations of } \tilde{\mathcal{A}}^k $;
		
		\State $\tilde{D}_{ii} \leftarrow \sum_j\tilde{\mathcal{A}}^{k-dilated}_{ij} $;
		
		\State $\hat{\mathcal{A}}^{k-dilated} \leftarrow \tilde{D}^{-\frac{1}{2}}\tilde{\mathcal{A}}^{k-dilated}\tilde{D}^{-\frac{1}{2}} $;
		
		\EndIf 
		
		\State $Z \leftarrow GCN(\hat{\mathcal{A}}^{k-dilated}, X')$;
		
		\State Calculate loss: $\mathcal{L} = - \sum_{l \in \mathcal{Y}_L } Y_l \cdot \ln (softmax(Z_l) )$;
		
		\State Perform backpropagation and update parameters;
		
		\EndFor 
		
		\State \textbf{return} $Z$.
		
	\end{algorithmic}
	
\end{algorithm}

\section{EXPERIMENTS}

\begin{table*}[]
	
	\caption{The meta-data of the three real-world datasets.}
	
	\label{dataset}
	
	\begin{tabular}{ccccccccc}
		
		\toprule
		
		dataset & edge(A-B) & number of A & number of B & number of A-B & training & validation & test & classes \\
		
		\midrule
		
		\multirow{3}{*}{DBLP} & Paper-Author & 14328 & 4057 & 19645 & \multirow{3}{*}{800} & \multirow{3}{*}{400} & \multirow{3}{*}{2857} & \multirow{3}{*}{4} \\
		
		& Paper-Conference & 14328 & 20 & 14328 & & & & \\
		
		& Paper-Term & 14328 & 8811 & 88420 & & & & \\ \midrule
		
		\multirow{2}{*}{IMDB} & Movie-Actor & 3015 & 4293 & 9041 & \multirow{2}{*}{800} & \multirow{2}{*}{400} & \multirow{2}{*}{1815} & \multirow{2}{*}{3} \\
		
		& Movie-Director & 3015 & 1676 & 3015 & & & & \\ \midrule
		
		\multirow{2}{*}{AMiner} & Paper-Scientist & 14209 & 4162 & 14422 & \multirow{2}{*}{800} & \multirow{2}{*}{400} & \multirow{2}{*}{2962} & \multirow{2}{*}{8} \\
		
		& Paper-Conference & 14209 & 2179 & 14209 & & & & \\
		
		\bottomrule
		
	\end{tabular}
	
\end{table*}

\subsection{Datasets.}

The details of the three real-world HINs in this work are shown in Table \ref{dataset}.

\textbf{$\bullet$ DBLP}. 
We extract a subset from DBLP, which contains 14328 ``Papers (P)"s, 4057 ``Authors (A)"s, 20 ``Conferences (C)"s, and 8811 ``Terms (T)"s \cite{wang2019heterogeneous}. The ``Terms (T)"s are processed as a feature of   ``Papers (P)"s. The ``Authors (A)"s have four classes: \emph{``Database"}, \emph{``Data Mining"}, \emph{``Information Retrieval"}, and \emph{``Machine Learning"}. For the meta-path-based baseline models, we employ the widely-used meta-path schemes \{APA, APCPA\}.

\textbf{$\bullet$ IMDB}\footnote{https://www.imdb.com/interfaces/}.
The IMDB is a dataset of movies. The experimental subset includes 3015 ``Movies (M)", 4293 Actors (A)", and 1676 ``Directors (D)". The ``Movies (M)"s have three classes: \emph{``Action"}, \emph{``Comedy"}, and \emph{``Drama"}. The widely-used meta-path schemes \{MAM, MDM\} are adopted in the meta-path-based models as baselines.

\textbf{$\bullet$ AMiner}\footnote{https://www.aminer.cn/citation}.
AMiner is also a computer science publication dataset. The subset involves 4162 ``Scientists (S)"s, 14209 ``Papers (P)"s, and 2179 ``Conferences (C)"s. In the node classification task, we have eight classes for ``Scientists (S)" : \emph{``computer scientists"}, \emph{``computational linguistics"}, \emph{``computer graphics"}, \emph{``computer networks \& wireless communication"}, \emph{``computer vision \& pattern recognition"}, \emph{``computing systems"}, \emph{``databases \& information systems"}, \emph{``human computer interaction"}, and \emph{``theoretical computer science"}. For the baseline models that use meta-paths, we employ \{SPS, SPCPS\}.

\begin{table*}[]
	
	\caption{The scores (\%) of Micro-F1 and Macro-F1 on the node classification task. The experiments do not perform the dilation.}
	
	\label{tab:classification}
	
	\begin{tabular}{cccccccc}
		
		\toprule 
		
		
		\multirow{2}{*}{method} & \multicolumn{2}{c}{DBLP} & \multicolumn{2}{c}{IMDB} & AMiner & \\
		
		& meta-path & Micro-F1/Macro-F1 & meta-path & Micro-F1/Macro-F1 & meta-path & Micro-F1/Macro-F1 \\ \hline
		
		\multirow{2}{*}{GCN} & APA & 49.84/47.00 & MAM & 58.95/42.50 & SPS & 25.93/24.18 \\
		
		& APCPA & 90.86/89.86 & MDM & 58.90/46.96 & SPCPS & 78.90/78.65 \\ \hline
		
		\multirow{2}{*}{GAT} & APA & 46.12/42.54 & MAM & 37.65/33.49 & SPS & 12.21/08.47 \\
		
		& APCPA & 71.93/71.20 & MDM & 40.23/35.03 & SPCPS & 44.98/38.50 \\ \hline
		
		HAN & APA+APCPA & 43.92/41.24 & MAM+MDM & 40.66/35.56 & SPS+SPCPS & 46.18/49.21 \\ \hline
		
		HCN & none & \textbf{91.49/90.75} & none & \textbf{64.79/55.87} & none & \textbf{88.25/88.19} \\
		
		
		\bottomrule
		
	\end{tabular}
	
\end{table*}

\begin{table*}[]
	
	\caption{The values (\%) of Normalized Mutual Information (NMI)  and  Adjusted Rand Index (ARI) on the node clustering task. }
	
	\label{tab:cluster}
	
	\begin{tabular}{ccccccc}
		
		\toprule
		
		
		\multirow{2}{*}{method} & \multicolumn{2}{c}{DBLP} & \multicolumn{2}{c}{IMDB} & \multicolumn{2}{c}{AMiner} \\
		
		& meta-path & NMI/ARI & meta-path & NMI/ARI & meta-path & NMI/ARI \\ \hline
		
		\multirow{2}{*}{GCN} & APA & 22.12/05.61 & MAM & 07.67/05.40 & SPS & 05.42/02.76 \\
		
		& APCPA & 68.88/74.01 & MDM & 10.64/09.24 & SPCPS & 46.39/33.91 \\ \cline{2-7} 
		
		\multirow{2}{*}{GAT} & APA & 22.12/05.90 & MAM & 07.78/05.81 & SPS & 02.09/00.95 \\
		
		& APCPA & 67.98/71.94 & MDM & 08.57/01.70 & SPCPS & 41.20/27.75 \\ \cline{1-7} 
		
		HAN & APA+APCPA & 66.40/72.96 & MAM+MDM & 11.26/09.98 & SPS+SPCPS & 28.03/16.11 \\ \hline
		
		HCN & none & \textbf{69.32/74.32} & none & \textbf{14.22/18.74} & none & \textbf{57.42/47.42} \\ 
		
		
		\bottomrule
		
	\end{tabular}
	
\end{table*}

\begin{figure}[h]
	
	\centering
	
	\includegraphics[width=0.6\linewidth]{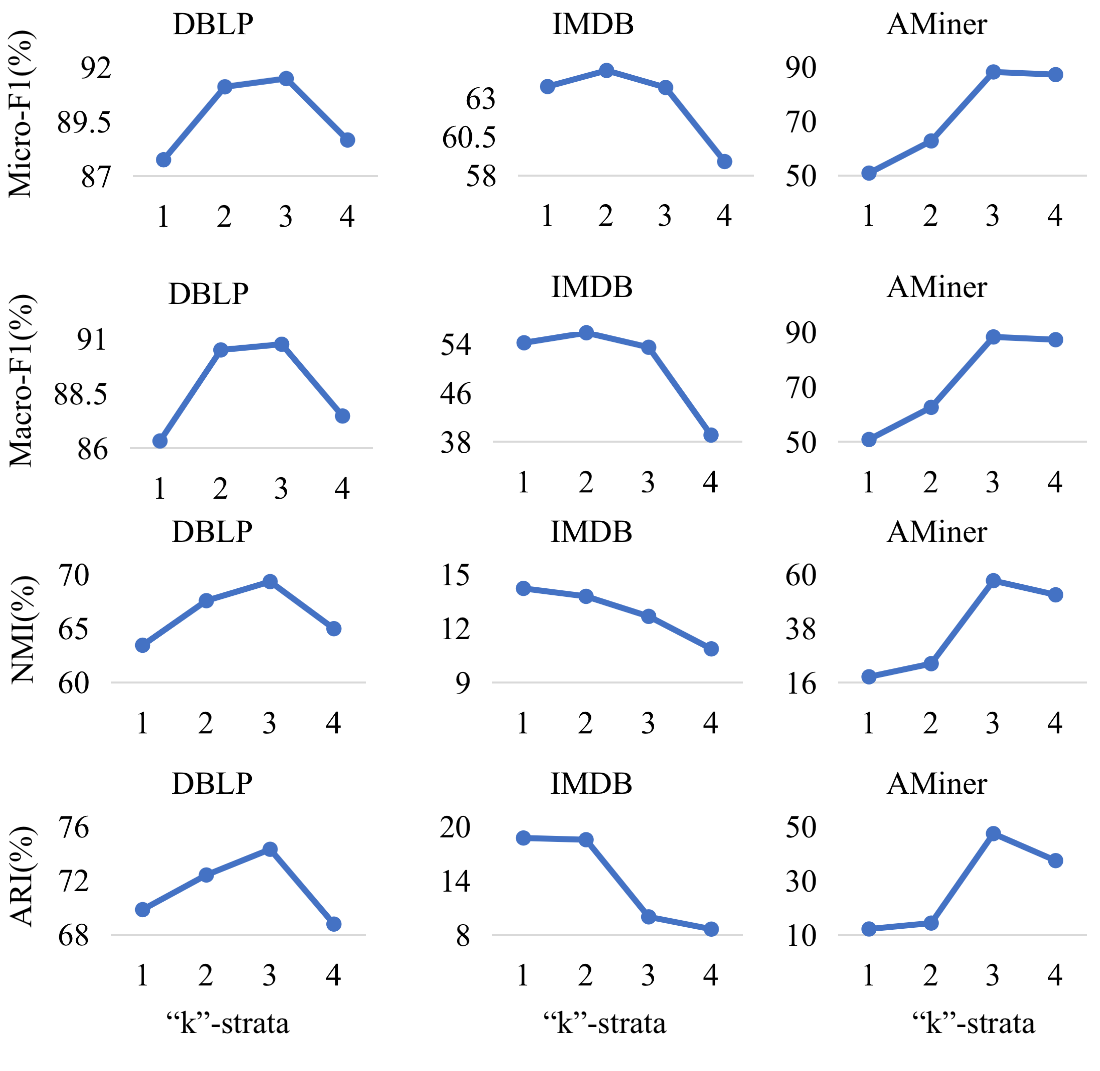}
	
	\caption{The values (\%) of Micro-F1, Macro-F1, Normalized Mutual Information (NMI), and Adjusted Rand Index (ARI) in different hyper-parameter $k$. As the $k$ increases, each curve achieves a peak and then drops. In classification, the 3-strata HCN achieves the highest Micro/Macro F1 scores in DBLP and AMiner; and the 2-strata HCN performs best in IMDB. In clustering, the 3-strata HCN achieves the highest NMI/ARI values in DBLP and AMiner; while the 1-stratum HCN performs best in IMDB. In each task, there is an optimal $k$ value and a balance in tuning $k$.  }
	
	\label{fig:khop}
	
\end{figure}

\begin{figure}[h]
	
	\centering
	
	\includegraphics[width=0.5\linewidth]{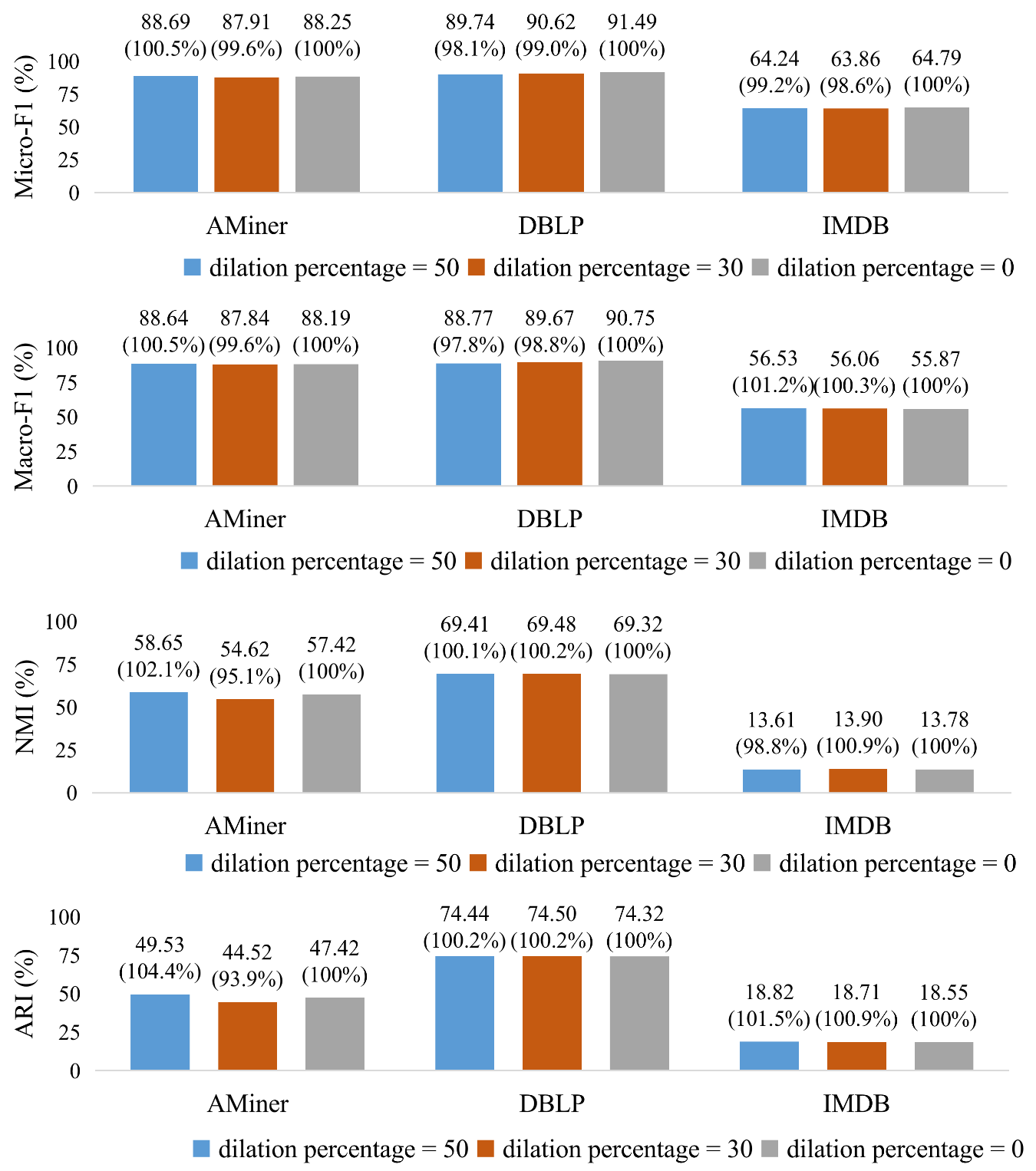}
	
	\caption{The results of the dilation. The X-axis represents the different dilation rates in AMiner, DBLP, and IMDB, respectively. The orange bars represent the results of a dilation percentage of $30$, which means we randomly drop 30\% and keep 70\% of the $k$-strata edges. The blue bars represent the results of a dilation percentage of $50$. The grey bars represent the results of no dilation, which correspond to the results in Table \ref{tab:classification} and Table \ref{tab:cluster} The Y-axis represents the values (\%) of Micro-F1, Macro-F1, Normalized Mutual Information (NMI), and Adjusted Rand Index (ARI), respectively. We present the absolute and relative values above the bars. We use the values of the grey bars as 100\%, and present relative values in percentage for the blue and orange bars. The ``mean $\pm$ standard deviation" of the relative values of the blue bars and the orange bars is $100.4\%\pm 1.8\%$ and $98.9\%\pm 2.2\%$, respectively. The $t$-tests find there is no statistical significance between the blue bars and the grey bars or between the orange bars and the grey bars. Both of the $p$-values are larger than 0.05. In brief, the dilation does not deteriorate the outcomes in the experiments.}
	
	\label{fig:k-strata-dilation}
	
\end{figure}

\subsection{Baselines}

The baseline models include three state-of-art models: GCN, GAT, and HAN.

\textbf{$\bullet$ GCN} \cite{kipf2016semi}. It is a semi-supervised graph convolutional network. Here we test all the aforementioned meta-paths in “5.1 Datasets” and report their performance respectively.

\textbf{$\bullet$ GAT} \cite{velivckovic2017graph}. It is a semi-supervised neural network that considers the attention mechanism on the homogeneous graphs. Here we test all the meta-paths.

\textbf{$\bullet$ HAN} \cite{wang2019heterogeneous}. HAN is composed of two parts: the GNN layers and a subsequent k-Nearest Neighbor (KNN) layer. The final outcomes of the node classification come from the KNN instead of GNN, although GNN is originally trained for the node classification task. The input of KNN is the output of the second-last layer in GNN. So HAN is not an end-to-end learning. Since GCN, GAT, and the proposed method do not use KNN, to perform a fair comparison, we keep the GNN layers but remove the KNN from HAN.

\textbf{$\bullet$ HCN}. The proposed meta-path-free representation learning for HNE. The codes will be open at Github.

\subsection{Implementation Details}

We stack two layers of GNN, which is commonly adopted in most GNNs. We randomly initialize parameters with uniform distribution. The optimizer of Adam \cite{kingma2014adam} and early stopping with the patience of 100 epochs are applied to update gradient. Besides, we set the learning rate to 0.01, the regularization parameter to 0.0005, the dropout rate to 0.5. 
The baseline methods use the same parameter setting. 

For DBLP and AMiner datasets, we set the number of hidden neurons to 64, while for IMDB the number of hidden neurons is set to 32. To ensure fairness, we split the datasets and use the same training, validation, and test set for all the models in this work.

\subsection{Multi-Class Classification}

To evaluate the performance of HNE, we perform multi-class classifications: four classes for ``Authors" in DBLP, three classes for ``Movies" in IMDB, and eight classes for `` Scientists" in AMiner. Please note that we do not perform the online dilation in the experiments of this section.

Table \ref{tab:classification} presents the results of Micro-F1 and Macro-F1 scores in the classification tasks. The proposed method performs best in all the three datasets. In detail, the 3-strata HCN achieves the highest scores of Micro-F1 and Macro-F1  in DBLP and AMiner; and the 2-strata HCN performs best in IMDB. 

The results also demonstrate that different meta-paths lead to different analytical outcomes. In DBLP, the $APCPA$ achieves much better classification results than $APA$ in both GCN and GAT; in IMDB, the ``MAM" and ``MDM" result in different results; and in AMiner, the ``SPCPS" results in better classification outcomes than ``SPS" in both GCN and GAT. 

Please note that the proposed HCN can achieve embedding of various node types, such as $A$, $P$, and $C$ in DBLP; $M$, $A$, and $D$ in IMDB; and $S$, $P$, and $C$ in AMiner. Comparatively, GCN, GAT, and HAN only learn embedding of one node type, such as $A$ in DBLP; $M$ in IMDB; $S$ in AMiner. 

\subsection{Node Clustering}

To further evaluate the performance of the HNE, we also perform clustering. We use K-means to cluster the nodes. The number of clusters is set to the number of classes in each dataset. Since the performance of K-means is influenced by initial centroids, all clustering experiments are conducted 10 times and the average results are reported. 

Table \ref{tab:cluster} summaries the clustering results under the metrics of Normalized Mutual Information (NMI) and Adjusted Rand Index (ARI) (\%).
The proposed method performs best in all the three datasets. In particular, the 3-strata HCN achieves the highest NMI/ARI values in DBLP and AMiner; while the 1-stratum HCN performs best in IMDB.

We find that the clustering results of different meta-paths are also different. The $APCPA$ achieves better results than $APA$ in DBLP; the ``MAM" and ``MDM" perform differently in IMDB; and the ``SPCPS" surpasses ``SPS" in AMiner. The results show researchers need to compare different meta-paths when using meta-path-based methods. 




\subsection{Hyper-Parameter $k$}


The experiments in this section evaluate how the hyper-parameter $k$ influences the performance. 
Figure \ref{fig:khop} illustrates the comparison of the different values of $k$. As the  $k$ increases, the curve in each subplot reaches a peak and then drops.
In other words, we find, in each task, there is an optimal $k$ value. 

The explanation could be as follows. In the beginning, as the hyper-parameter $k$ increases, more composite relations are generated, which contribute to better analytical outcomes. 
Take the DBLP-like network in Figure \ref{fig:k_strata} as an example. 
For the given node $A_1$, when $k$ becomes $2$, new two-hop composite relations, such as $A_1P_1A_2$ (a co-authorship between two authors) and $A_1P_1C_2$ (a participation relation between an author and a conference), capture more semantics and therefore improve analytical outcomes. 
Nonetheless, when the $k$ is too big,  the $k$-hop composite relations with long distances may bring in weak relations and even noises, which damage the analytical outcomes. 


In conclusion, we need to tune an appropriate number of $k$. 
The integer $k$ is a hyper-parameter, just as the number of layers or neurons in a fully-connected neural network. One can use the grid search to find the optimal $k$ automatically.


\subsection{Online Dilation}


%

%

%


Figure \ref{fig:k-strata-dilation} evaluates the results of the online dilation. 
The X-axis represents the different dilation rates in AMiner, DBLP, and IMDB, respectively. 
The orange bars represent the results of a dilation percentage of $30$, which means we randomly drop 30\% and keep 70\% of the $k$-strata edges. 
The blue bars represent the results of a dilation percentage of $50$. 
The grey bars represent the results of no dilation, which correspond to the results in Table \ref{tab:classification} and Table \ref{tab:cluster} 
The Y-axis represents the values (\%) of Micro-F1, Macro-F1, NMI and ARI, respectively. 
We present the absolute and relative values above the bars. 
We set the values of the grey bars to 100\%, and calculate relative values in percentage for the blue and orange bars. 
The ``mean $\pm$ standard deviation" of the relative values of the blue bars and the orange bars in all the experiments in Figure \ref{fig:k-strata-dilation} is $100.4\%\pm 1.8\%$ and $98.9\%\pm 2.2\%$ , respectively. 
By $t$-tests, we find these values have no statistical significance,
which means dropping 30\% or even half of $k$-strata edges does not make the analytical outcomes worse.

Why the dilation does not damage the analytical results? Take the DBLP-like network in Figure\ref{fig:k_strata} as an example. There are 10 two-strata edges that connect to $A1$: $A1-P1$, $A1-P2$, $A1-P3$, $A1-P4$, $A1-P5$, $A1-A2$, $A1-A3$, $A1-A4$, $A1-C1$, and $A1-C2$, as Figure\ref{fig:k_strata_adj} shows. If the dilation percentage is 30,  we randomly drop 3 edges such as $A1-P3$, $A1-P4$, and $A1-P5$, and remain the left 7 edges including $A1-A4$. By the consecutive relations of $A1-A4-P3$, $A1-A4-P4$, and $A1-A4-P5$ in the $4$-strata adjacency matrix, $A1$ can still extract information from $P3$, $P4$, and $P5$ through a GNN. 

The ``online dilation" conducts a different random drop in a few epochs. For one thing, the whole training process does not lose any information since the dilation performs different random drops in epochs. For another, in theory, the ``online dilation" incorporates more diversity into the input data and therefore prevents over-fitting and reduces message passing \cite{rong2019truly}. In practice, although we do not find statistically-significant improvements in this work after we adopt the dilation rate of 50\% or 30\%, dropping even a half of edges does not deteriorate the analytical results but make the $k$-strata adjacency matrix sparse. In real-world projects when we need to embed huge knowledge graphs, the dilation is supposed to save training costs without sacrificing accuracy.


\section{CONCLUSION}

In this work, we propose a novel meta-path-free representation learning on a HIN. The proposed method overcomes the challenge of heterogeneity and captures both the semantic and structural information. The experimental results demonstrate that the proposed method significantly outperforms the state-of-the-art methods in the various tasks. Hopefully, this work can inspire more researches on meta-path-free HNE.


\bibliographystyle{unsrt}  
\bibliography{ref}

\end{document}